\documentclass[3p]{elsarticle}

\usepackage{amssymb}
\usepackage{amsmath}
\usepackage{lineno}
\usepackage[dvipsnames]{xcolor}

\journal{Applied Mathematics and Computation}

\begin{document}


\begin{frontmatter}

\title{Suppressing defection by increasing temptation: the impact of smart cooperators on a social dilemma situation}

\author[label1,label2]{Hsuan-Wei Lee}
\cortext[mail1]{Email addresses: hwwaynelee@gate.sinica.edu.tw; szolnoki.attila@ek.hun-ren.hu}
\author[label3]{Colin Cleveland}
\author[label4]{and Attila Szolnoki}

\address[label1]{Institute of Sociology, Academia Sinica, Taiwan}
\address[label2]{Department of Physics, National Taiwan University, Taiwan}
\address[label3]{Department of Informatics, King's College London, London, UK}
\address[label4]{Institute of Technical Physics and Materials Science, Centre for Energy Research, P.O. Box 49, H-1525 Budapest, Hungary}

\begin{abstract}
In a social dilemma situation, where individual and collective interests are in conflict, it sounds a reasonable assumption that the presence of super or smart players, who simultaneously punish defection and reward cooperation without allowing exploitation, could solve the basic problem. The behavior of such a multi-strategy system, however, is more subtle than it is firstly anticipated. When exploring the complete parameter space we find that the emergence of cyclic dominance among strategies is rather common, which results in several counter-intuitive phenomena. For example, the defection level can be lowered at higher temptation, or weaker punishment provides better conditions for smart players. Our study indicates that smart cooperators can unexpectedly thrive under high temptation, emphasizing the complexity of strategic interactions. This study suggests that the principles governing these interactions can be applied to other moral behaviors, such as truth-telling and honesty, providing valuable insights for future research in multi-agent systems.

\end{abstract}


\end{frontmatter}

\section{Introduction}

To understand why we cooperate while defection offers short-term benefits is the central problem not only in social science, but also in biology, ecology, or economics~\cite{perc_pr17,ohdaira_srep22,fu_y_c21,fu_mj_pa19,quan_j_pa21}. Because of its importance, as a natural reaction from research community, it has collected significant scientific interest from the mentioned fields~\cite{xiao_sl_epjb22,ye_yx_epjb22,lv_r_csf24,javarone_jsm24,zhang_y_amc24,lv_r_csf23,ding_r_csf23}. Evidently, there is no a single answer and over the last decades several mechanisms or external conditions have been identified which support the emergence of cooperation~\cite{kang_hw_csf24,zhang_h_csf23,han_j_c24,duan_yx_csf23,bin_l_amc23,lu_sn_csf23,liao_hm_amc23,szolnoki_amc20}. A quite straightforward idea could be to directly support those who cooperate or punish the rival participants hence establishing competitive payoff situation for good players~\cite{szolnoki_epl10,ohdaira_csf24,feng_sn_pa24,szolnoki_pre11b,lv_csf23,hua_sj_csf23}. However, the devil is in the details because there is no free lunch and maintaining the abovementioned incentives requires extra care, hence extra cost from the supporting group~\cite{ohdaira_plr23,xiao_jf_jpc23,li_k_csf21,quan_j_c19,lee_hw_csf24}. This often makes these institutions vulnerable by generating a second-order free riding dilemma where simple cooperators enjoy the benefits of these incentives without contributing to the related expenses~\cite{shen_y_csf23,sun_xp_amc23,wang_xj_csf22,hua_sj_eswa24}.

Our present study targets exactly this situation where we extend the traditional two-strategy prisoner's dilemma model by adding a new strategy who has some extra capacity and skill. In particular, these players, referred to as smart cooperators, not only cooperate with partners but also are capable to reward those partners who also cooperate. Furthermore they recognize defection, hence they avoid being exploited and punish defectors partners simultaneously. On the other hand, the model becomes realistic only if we assume that these smart players should bear and extra permanent cost, which makes it possible to ``cover'' the incentives and their extra information to avoid being exploited. The competition of these strategies is studied within the framework of prisoner's dilemma game which is capable to grab the essence of the basic dilemma we have mentioned above. 

Despite the simplicity of this social game and the applied strategies we still have several parameters, including reward, punishment, temptation, and extra cost of smart players, which offer a huger liberty how system may evolve. To explore these scenarios we have systematically scan the whole parameter space and identify the stationary fractions of competing strategies. Our key questions include whether the presence of smart players can always drive the system toward the expected direction, or is it always beneficial to apply large punishment? Or, is it always good for smart players if their extra cost is minimal? These questions seem to be straightforward, but as we demonstrate below, the answers are more delicate. Moreover, despite the simplicity of our model it is capable to provide the general system behavior that is frequently observed when multi strategies are present in the fight for space. 

The rest of this paper is organized as follows. In Section~\ref{model}, we define the details of our model and specify the methodology of our computational experiments. It is followed by consequent results and observations which are presented in Section~\ref{result}. Last, our conclusions and some general comments are given in Section~\ref{conclusion}. 

\section{Model}
\label{model}

Our spatial model is an extension of traditional prisoner's dilemma game where unconditional defector ($D$) and unconditional cooperator ($C$) players compete for space on an $L \times L$ square lattice where periodic boundary conditions are applied. In agreement with previous studies, in the payoff matrix elements we fix the reward for mutual cooperation as $R=1$, and the punishment for mutual defection $P=0$, while the remaining two elements, such as the $T$ temptation to defect and $S$ sucker's payoff are free parameters~\cite{du_cp_amc23,kang_hw_pla21,li_xy_epjb21,liu_rr_amc19}. The third strategy represents a sort of ``smart'' cooperator ($M$) who cooperates with other cooperators, but is also capable to recognize defectors. Consequently, when an $M$ player meets with a $D$ player then the latter remains empty handed and the former avoids being exploited. Furthermore, an $M$ player applies both positive and negative incentives to support cooperation. Therefore cooperator players are awarded by a reward $W$ and defector players are punished by $U$ when they face with an $M$ partner. Evidently, to maintain the incentives and to monitor the behavior of neighbors require extra efforts from smart cooperators, which is considered via an additional $G$ cost they have to bear. In this way their advantage, or their positive impact on general cooperation is less straightforward. The relations of different strategies are summarized in the following payoff matrix:
\begin{center}
	\begin{tabular}{r|c c c}
		& \,\,\,\,\,\,$M$ & $C$ & $D$\\
		\hline
		$M$ &\,\,\,\,\,1+$W$-$G$ & \,\,\,\,\,1-$G$ &-$G$\\
		$C$ & 1+$W$ & 1 & \,\,$S$\\
		$D$ & \,\,$-U$ & $T$ & \,\,0\\
	\end{tabular}\,\,.
\end{center}
It is worth noting that a slightly similar payoff parametrization was suggested recently by Zhao {\it et. al} in Ref.~\cite{zhao_amc23}. But in their model a defector can still utilize a connection with a smart player. Furthermore, the latter strategy is also rewarded by an unconditional cooperator which is unjustified in the absence of additional resources. We strongly believe that our payoff matrix, shown above, expresses the role of strategy $M$ more accurately: these players possess the requested resources to motivate cooperation for an additional price they have to bear.

When the strategy evolution is monitored we follow the standard protocol. More precisely, we randomly select a player $i$ and a neighboring player $j$. If their strategy $s_i$ and $s_j$ are different then we calculate their $\Pi_{i}$ and $\Pi_{j}$ payoff values by considering their nearest neighbor links to their neighbors. After player $i$ adopts the $s_j$ strategy of player $j$ with the probability
\begin{equation}
	\Gamma(s_j \to s_i)=\frac{1}{1+\exp[(\Pi_{i}-\Pi_{j}) /K]}\,\,.
	\label{fermi}
\end{equation}
Here, we apply the broadly used $K=0.1$ noise parameter value that generally allows player $i$ to imitate the more successful strategy, but seemingly irrational strategy choice is also possible with a low probability~\cite{szabo_pre05,deng_ys_pa21,liu_jz_epjb21,lv_amc22,yu_fy_csf22,quan_j_csf21}. This process is repeated $N = L \times L$ times which establishes a full Monte Carlo (MC) step. Typically we use $10^4$ MC steps to reach the stationary state where the linear system size is $L=1000$ or $L=1200$. There is, however, an additional technical detail which has been proven to be very useful to obtain reliable numerical data. In particular, we apply a very small $\mu = 10^{-6}$ mutation rate which allows players to update their strategy randomly. It technically means that one in a million elementary updating process is random where the imitation probability defined by Eq.~\ref{fermi} is ignored. This tiny mutation rate helps us to avoid very heavy finite-size effects without influencing the pattern formation relevantly. Evidently, the usage of mutation prevents the system from being trapped in a homogeneous state, hence we cannot talk ``phases'' anymore, but a strategy is not really viable if its fraction is proportional to the mutation rate.

In the rest of this work we fix the sucker's payoff $S=-0.1$ which means a proper prisoner's dilemma situation. According to the payoff matrix, four other parameters remain that characterize the $T$ temptation to defect, the $W$ extra reward to support cooperation, the $U$ punishment for defection, and the $G$ extra cost of smart players to maintain the incentives. In the following we explore the full parameter space and recognize the typical system behavior thanks to the presence of smart cooperator players.

\section{Results}
\label{result}

\begin{figure}[h!]
	\centering
	\includegraphics[width=0.45\textwidth]{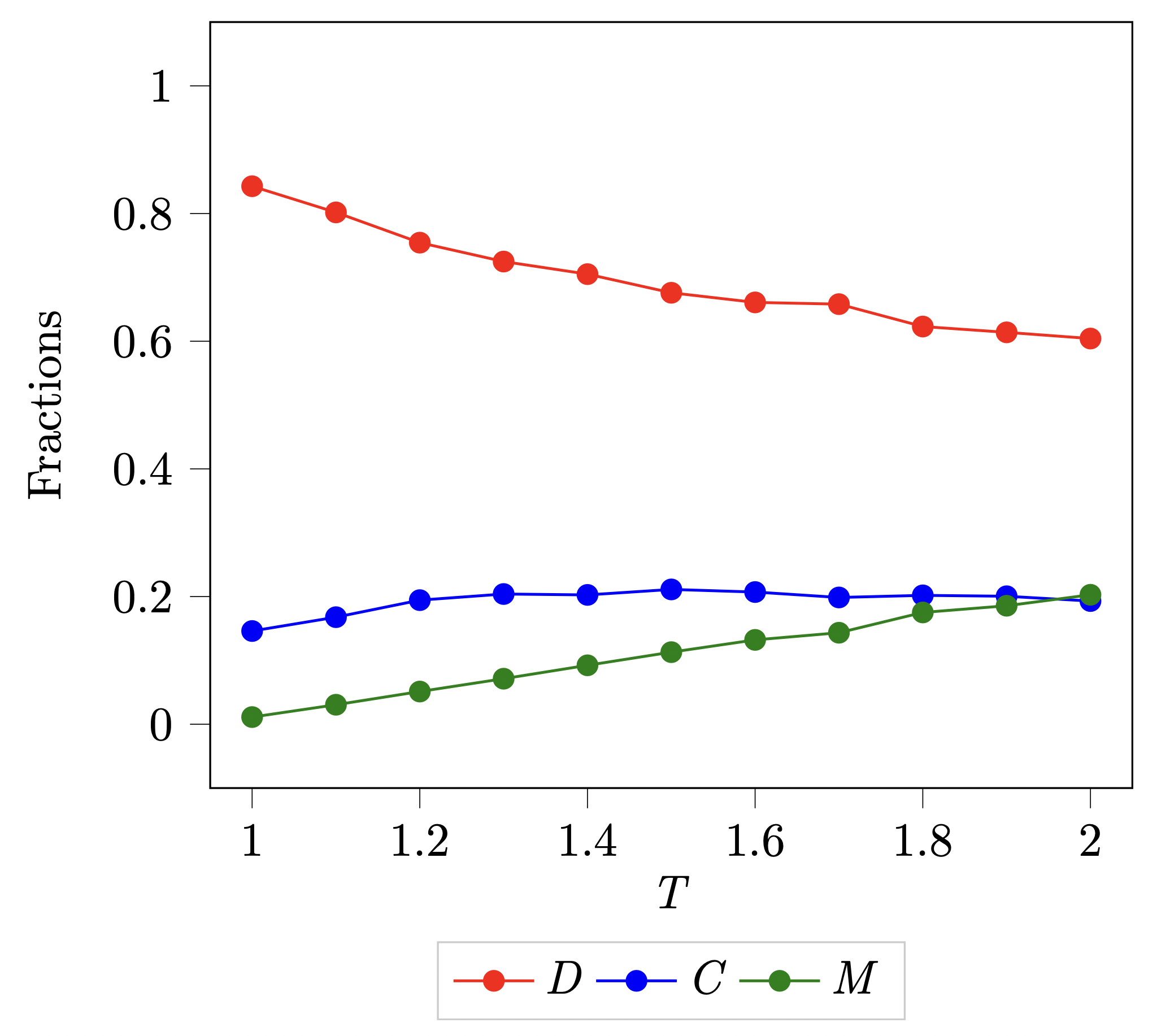}	
	\caption{The stationary fractions of competing strategies in dependence of $T$. Interestingly, the portion of defectors decays as we increase the temptation value. Other parameters are $U=1$, $W=0.5$, $G=0.6$, and $S=-0.1$.}
	\label{motiv}
\end{figure}

But first, as a motivation plot, we present the fractions of all strategies at representative parameter values, which illustrate nicely how unexpected the consequence of an advanced strategy could be. This is shown in Fig.~\ref{motiv} where we plot the stationary fractions of strategies in dependence of temptation level. At $T=1$ $M$ players practically die out due to the large $G$ cost value they have to sustain. More precisely, their fraction is so small that is proportional to the mutation rate. As a consequence, defectors dominate the system. The na{\"i}ve expectation would be that this situation becomes even worse if we increase the temptation value because it typifies a direct form of support to defector strategy. Interestingly, however, the opposite happens and the portion of defectors decays gradually by increasing $T$. In parallel, the fraction of $M$ players increases significantly despite of the aforementioned large cost level.

\begin{figure}[h!]
	\centering
	\includegraphics[width=0.45\textwidth]{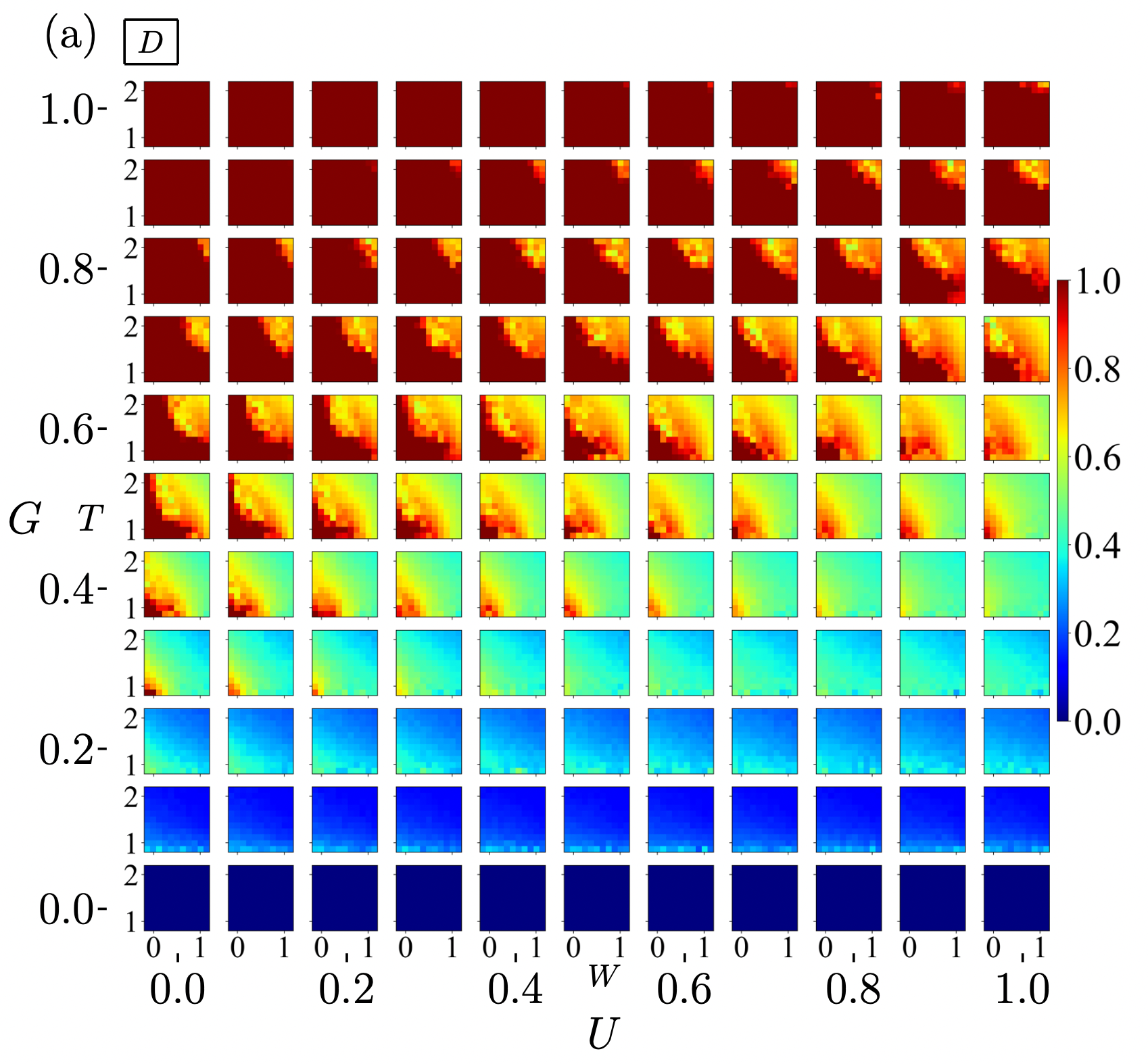}	
	\includegraphics[width=0.456\textwidth]{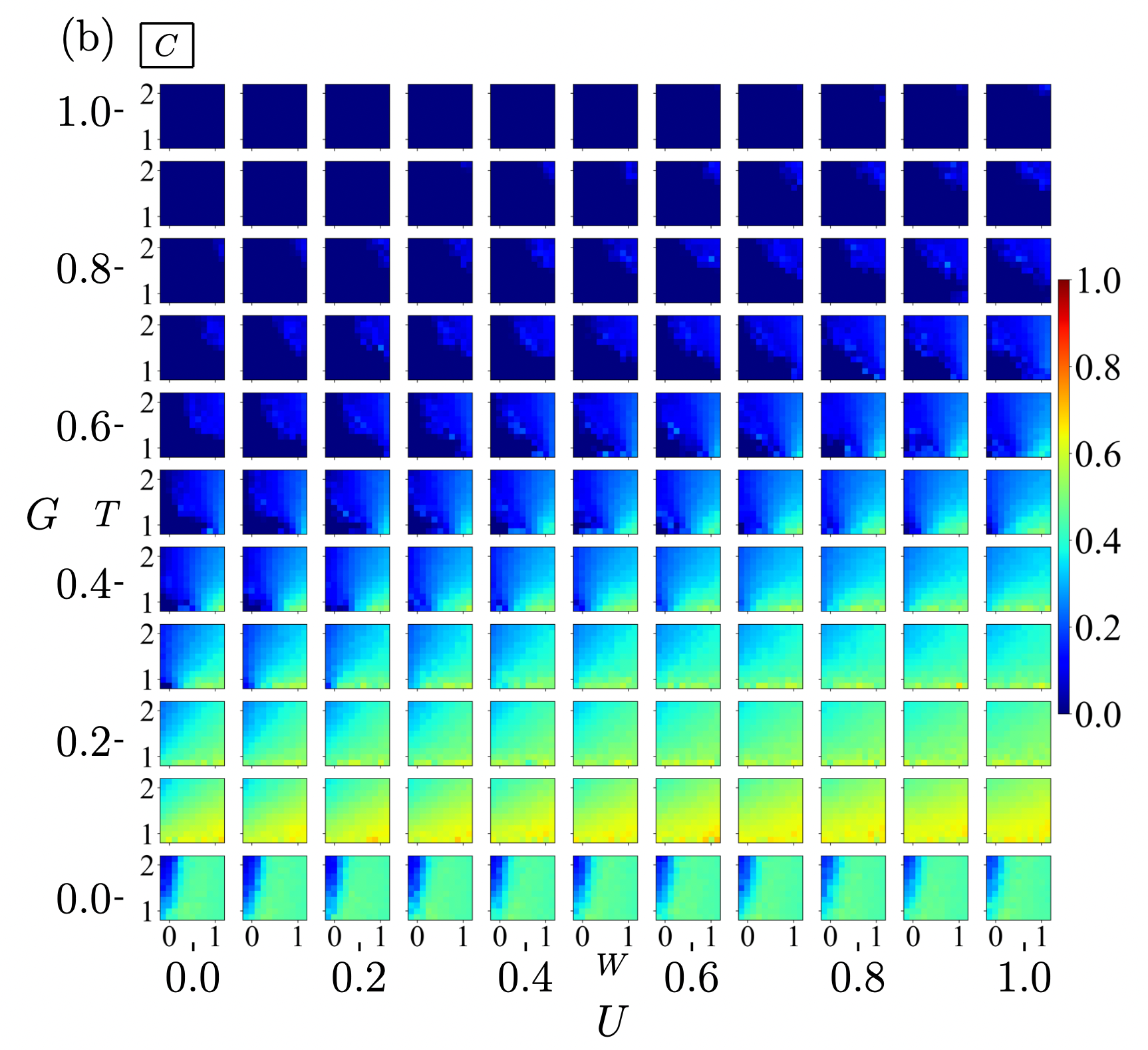}\\	
	\includegraphics[width=0.45\textwidth]{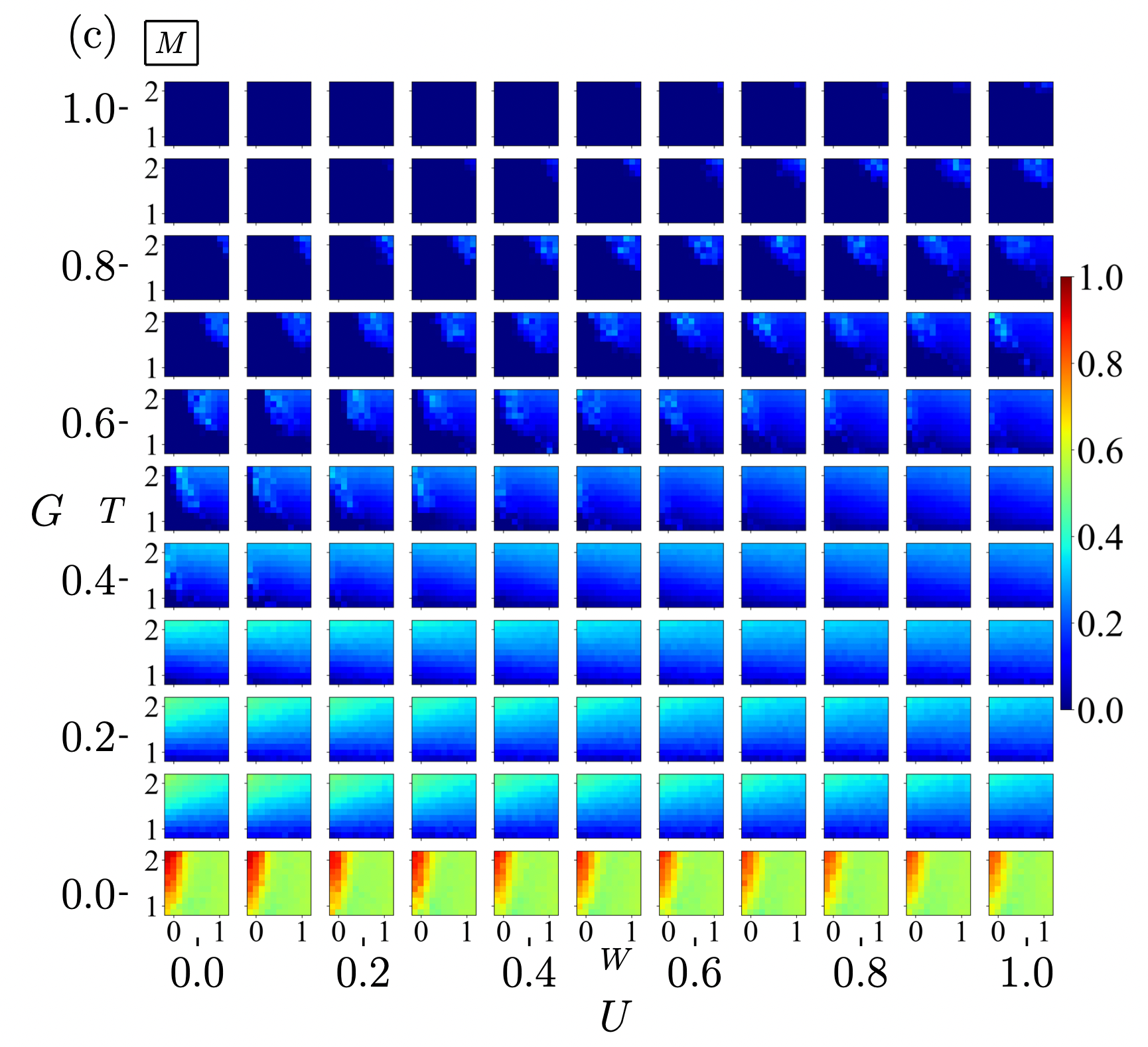}	
	\caption{Heat map of the stationary fraction of $D$ (top-left), $C$ (top-right), and $M$ (below-central) strategy on the full $T-W$ parameter plane. Each micro panel was obtained at a fix value of $G$ and $U$, as indicated on the axis. The value of $S=-0.1$ was fixed for all cases.}
	\label{plane}
\end{figure}

The above-discussed phenomenon, as we explain it later, is not necessarily counter-intuitive if we consider the relation of competing strategies more carefully. But before doing so we first give a general overview about the system behavior in the vast sea of parameters. It can be done systematically if we present the general level of a specific strategy in a $T-W$ parameter plane at fixed $G$ and $U$ values. After we modify the values of the latter parameters systematically in a reasonable interval. The repeat of this loop results in $11^2 = 121$ different heat maps for each strategy. We stress again that traditional phase diagrams cannot be presented because of the nonzero mutation rate, hence we need to present the portions of all three strategies separately. The results are shown in Fig.~\ref{plane}.

By checking these panels some general observations can be made. If the $G$ additional cost of the $M$ strategy is too high, then smart players are not really functioning and the traditional relation between $C$ and $D$ players results in a complete dominance of the latter strategy almost everywhere. More precisely, the only exception is when $U$ punishment is high enough , thereby availing a chance for $M$ strategy at the expense of defectors. This phenomenon is more pronounced at higher temptation values; hence, the phenomenon we presented in Fig.~\ref{motiv} is a general effect in a broad range of parameters. When $G$ is small, then defection can be kept at a minimal level and the real victors of this process are unconditional cooperators. Interestingly, smart player can dominate the population only at very small $G$ cost values, but only if the $W$ reward remains below a threshold level, otherwise it is better for a player to choose the unconditional cooperation strategy. The latter effect is almost independent of the $U$ punishment level, but it is straightforward because $D$ players are present just minimally at these parameter values. However, the threshold value of $W$, where $C$ players can beat $M$ players, depends sensitively on the $T$ temptation value, as the low row illustrates in Fig.~\ref{plane}c. The only reason why this fact is worth noting is because the payoff elements characterizing the relation of $C$ and $M$ strategies do not contain parameter $T$. Nevertheless, this observation gives a clue to understand the above described phenomena more deeply.

\begin{figure}[h!]
	\centering
	\includegraphics[width=1.0\textwidth]{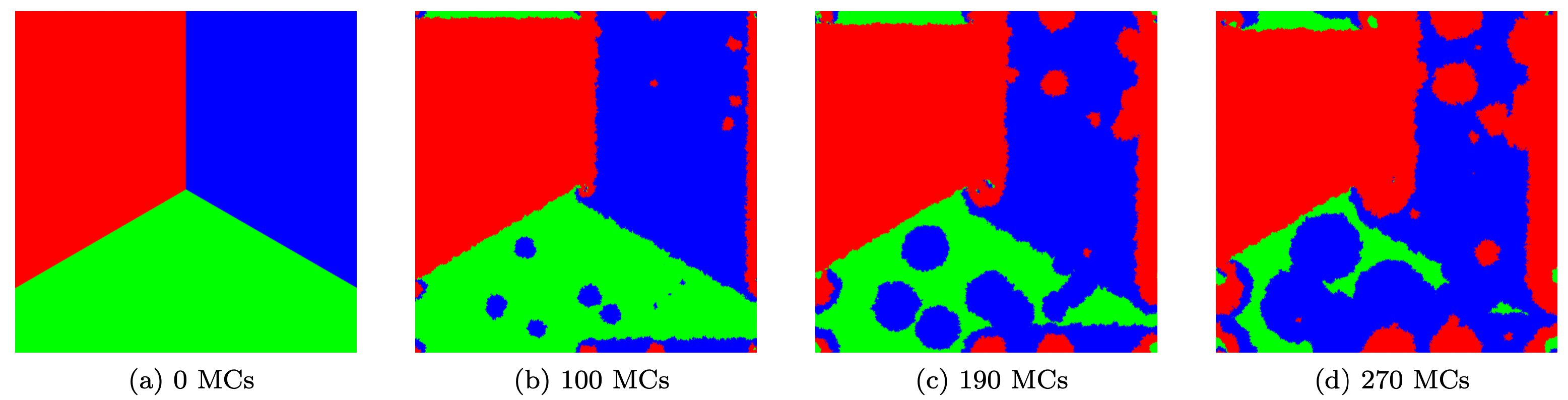}	
	\caption{Propagation of invasion fronts starting from a prepared initial state, shown in panel~(a). Color coded identically to Fig.~\ref{motiv}. Blue cooperators invade the green bulk of smart cooperators, while red defectors invade blue cooperators. To close the loop, the green domain gently intrudes into the red field. Time in MC steps are shown below. Parameter values are $T=1.6$, $W=0.3$, $G=0.5$, $U=0.2$, and $S=-0.1$.}
	\label{cyclic}
\end{figure}

These seemingly paradoxical effects can be explained if we recognize the potential intransitive relations of competing strategies. Evidently, defectors beat unconditional cooperators because of the negative $S$ sucker's payoff value. However, $C$, outperforms $M$ because the former enjoys being rewarded by $W$ without paying its cost $G$. Last, $M$ can exceed $D$ because the former punishes the latter and avoids being exploited. This closes the loop and establishes a rock-scissors-paper-like cyclic dominance among the three strategies. To support this conjecture, we present a front propagation process between homogeneous domains in Fig.~\ref{cyclic}.
 
As Figure~\ref{cyclic}a shows, we launch the evolution from a prepared initial state where homogeneous domains of competing strategies are arranged. They are red defectors, blue cooperators, and green smart players. The subsequent panels show the stages of front propagation after different MC steps, as indicated in the label. Already, panel~(b) illustrates the key effects that confirm our basic conjecture. Namely, the border separating the red and blue areas starts moving to the right, indicating that $D$ players start invading $C$ players. Simultaneously, there is a certain chance that the seed of defectors emerge in the bulk of unconditional cooperators due to mutation and these ``small red domains" start growing. Hence, we can conclude that $D$ beats $C$, as expected. Of course, green $M$ players can also emerge due to mutation in the blue sea, but they cannot spread because they are inferior to $C$ players. Actually, the opposite is true, as the shift of border between green and blue domains indicates. Furthermore, a small seed of blue payers can easily grow in the green ocean. Both movements underline the superiority of $C$ against $M$. Finally, the green domain can invade the red domain of defectors. This process is weaker than the previously discussed cases, because the $M \to D$ invasion is less effective, but it still exists, as it can be seen clearly on the top part of the (b)-(d) panels. In sum, the movement of domain walls confirm the $D \to C \to M \to D$ intransitive relation undoubtedly.

\begin{figure}[h!]
	\centering
	\includegraphics[width=0.45\textwidth]{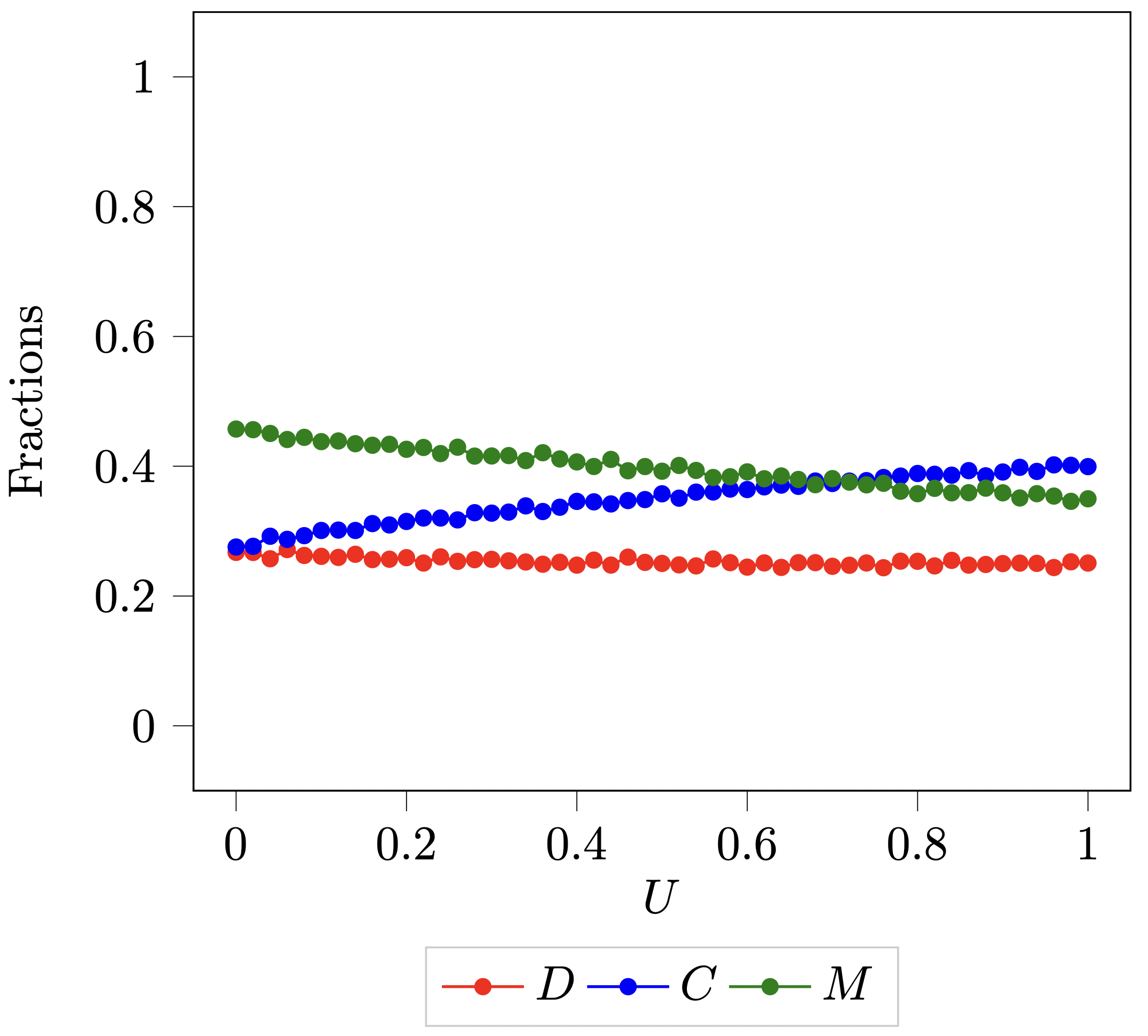} 	
	\caption{The stationary fractions of competing strategies in dependence of $U$. Interestingly, smart cooperators benefit the most from low punishment level. The increase of punishment, however, just slightly modify the defection level. Instead, cooperators enjoy it at the expense of smart cooperators. Other parameters are $W=0.2$, $G=0.2$, $T=2$, and $S=-0.1$.}
	\label{punish}
\end{figure}

Based on our knowledge about cyclically dominant systems~\cite{szolnoki_epl20,avelino_epl21,szolnoki_prr24,baker_jtb20}, the above-described counter-intuitive behaviors can be understood in the following way. It is a frequently observed and broadly confirmed effect that when we support of species in a loop directly, then its predator will benefit from this intervention mostly~\cite{tainaka_pla95,szolnoki_srep23,yang_csf23,avelino_pre19b}. Exactly the same system behavior can be seen in our motivation figure, as shown in Fig.~\ref{motiv}: when we increase the $T$ temptation level and seemingly support defectors, then their ``predators',' namely, smart players will benefit from this action. According to this argument, we can expect other apparently unexpected system behaviors. For example, it is expected that the increase of punishment $U$ will weaken defectors. Indeed, it is true in the high or significant $G$ cost region where the role of smart players is negligible. When the latter strategy is viral, however, we can see that by weakening $D$, we indirectly penalize its ``predator'' that is the mentioned strategy $M$. This effect can be seen in Fig.~\ref{punish}, where we plot the fractions of strategies in dependence of parameter $U$. The figure suggests that the growth of $U$ just slightly reduces the portion of $D$ and its predator $M$ strategy suffers more. However, the ``prey'' of strategy $D$, which is unconditional $C$, benefits from harsh punishment. As $M$ is the predator, while $C$ is the prey of strategy $D$, it explains why the former two cooperating strategies behave differently by increasing punishment exclusively.

\begin{figure}[h!]
	\centering
	\includegraphics[width=0.45\textwidth]{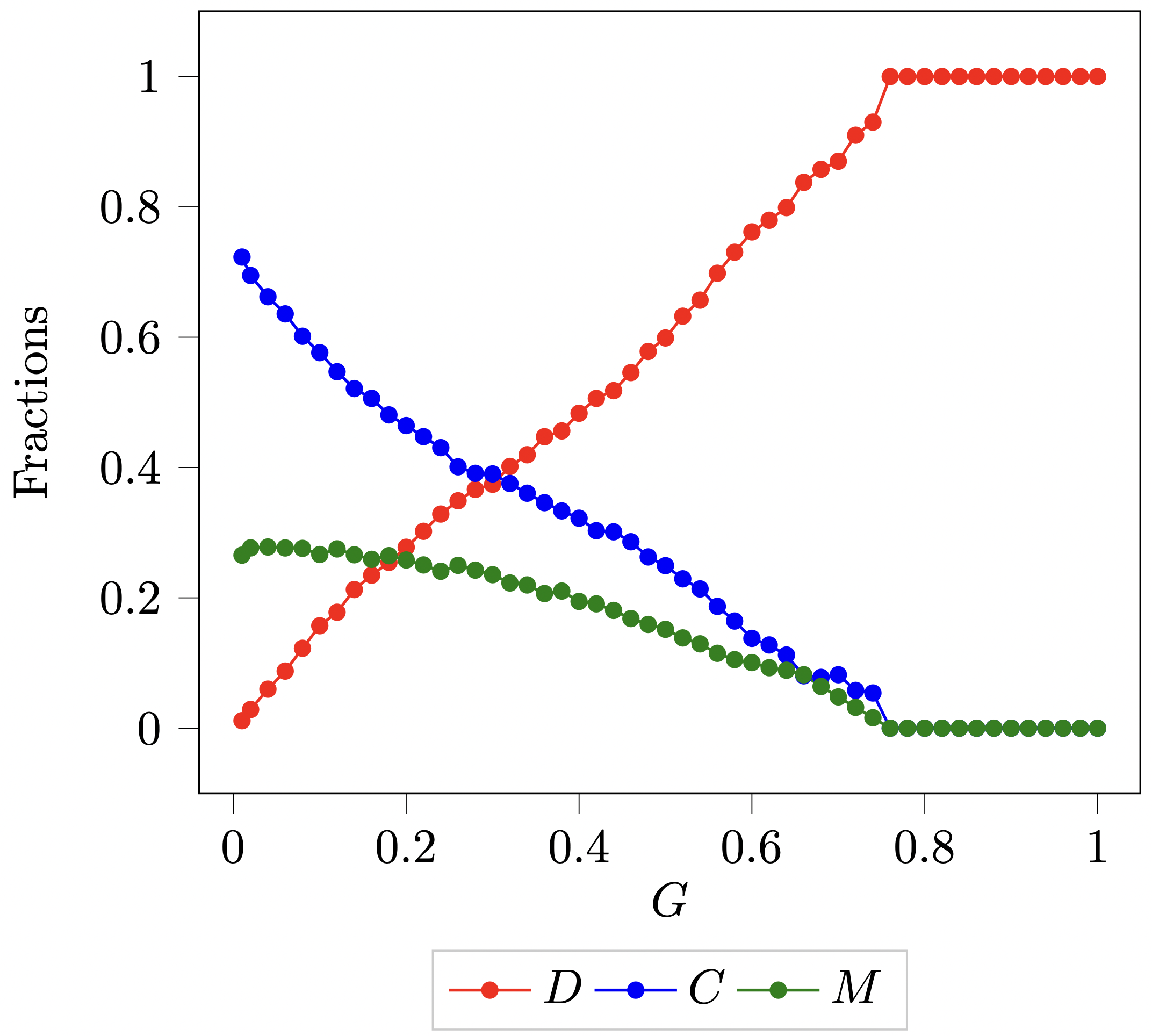}
	\caption{The stationary fractions of competing strategies in the dependence of $G$. Interestingly, simple cooperators benefit the most from the low cost of smart cooperation and the increase of this parameter value smites strategy $C$ more drastically. Other parameters are $W=0.6$, $U=0.6$, $T=1.5$, and $S=-0.1$.}
	\label{cost}
\end{figure}

It also seems a straightforward expectation that lowering the $G$ cost of the $M$ strategy should directly support smart players. Especially because the mentioned parameter is present only in the top role of payoff matrix that determine the incomes of strategy $M$. Indeed, it is generally true, but the real beneficiary of the mentioned intervention should be $M$'s predator, which is an unconditional cooperator. Figure~\ref{cost} confirms our ``advanced expectation'' because the portion of strategy $C$ increases more drastically when we lower the cost value $G$ of smart players.

\section{Conclusion}
\label{conclusion}

One of the typical features of complex systems is an unexpected scenario in response to external interventions~\cite{san-miguel_fcs23}. Evidently, when there are more than two options for participants to become successful in a multi-agent population, then the mentioned complexity is almost an inevitable ingredient of the collective behavior~\cite{sun_xp_pla23,li_my_csf22,liu_lj_rspa22,wei_x_epjb21,flores_jtb21}. By introducing a seemingly trivial strategy into the intensively studied enigma where unconditional strategies compete~\cite{wang_sx_pla21,szolnoki_epl15,xie_k_csf23,zhu_pc_epjb21}, our principal motivation is to check whether the increment of the number of strategies may result in a system response that cannot be predicted in advance. What can we expect from a ``super player'' who is armed with a specific skill to recognize the strategy of partners? Hence, it is protected against being exploited? On top of that, it can directly support positive acts and retaliate against negative ones and thus can modify competitors' incomes. Evidently, these players are expected to be the champions of an evolutionary process. To make the model more reliable, however, we assume that they have to pay the price of this superpower. 

Our first key observation is that this super strategy is generally unable to prevail and $M$ players are rarely dominant. More precisely, they can only win when their additional cost is extremely small and the reward they provide for cooperation is also moderate. Notably, the level of punishment and how they penalize defection have no particular role in their final dominance. In all other cases, the real benefit of their presence is unconditional cooperators who can be considered as second-order free riders. At this point, it is worth noting that the very low mutation rate, which does not affect the pattern formation relevantly, but helps to avoid serious finite-size problems, has an additional positive consequence: second-order free riders cannot repair the original dilemma situation that is frequently reported in other models where cooperation supporting incentives are introduced~\cite{panchanathan_n04}.

The explanation as to why smart players become just one of the options is the fact that their presence introduces a cyclical dominance among competing strategies. In a vast variety of full parameter space, the $M$ strategy beats $D$ who beats $C$ who beats $M$, hence closing the loop. Importantly, a similar emergence of intransitive relations among competing strategies has been reported quite often in social dilemma situations~\cite{liu_lj_rsif22,flores_amc24,lee_hw_amc22,szolnoki_csf20,cheng_f_amc20,tao_yw_epl21,szolnoki_njp14,yang_hx_epl20,szolnoki_pre10b,gao_sp_pre20}.

Having realized the above-described relations, the seemingly counter-intuitive system behavior becomes understandable immediately. We already knew for more than two decades that a direct aid for a loop member will mostly support its predator partner, and similarly, the main loser of a weakening process will be the target's superior fellow. Accordingly, increasing the temptation to defect can result in the blooming of smart players in our present model. Or, the latter strategy would suffer more by increasing the severity of punishment. 

We also recognize that the principles explored in this study could extend beyond cooperation to other aspects of moral behavior, such as truth-telling or honesty. Recent reviews, such as \cite{capraro2021mathematical}, provide a foundational basis for this line of inquiry. Future research could explore additional strategies, different types of social dilemmas, and the application of our findings to other aspects of moral behavior, such as truth-telling and honesty. These directions offer valuable opportunities to further understand strategic interactions in multi-agent systems. The main message of our present study is that when we introduce more strategies to interpret a real-life situation more faithfully, it always gives a chance for a loop to emerge among competitors. Therefore, the consequence of a new rival on global behavior will not necessarily be our first na{\"i}ve expectation; hence, the external conditions that can determine the payoff elements of participants should be designed with special care.

\vspace{0.5cm}
This research was supported by the National Research, Development and Innovation Office (NKFIH) under Grant No. K142948 and by the National Science and Technology Council of the Republic of China (Taiwan) under Grant No. NSTC112-2410-H-001-026-MY2.

\bibliographystyle{elsarticle-num-names}

\end{document}